%% file: main.tex
\documentclass[fleqn,10pt]{wlscirep}
\usepackage[utf8]{inputenc}
\usepackage[T1]{fontenc}
\usepackage{layout}
\usepackage{preamble}
\title{Towards pandemic preparedness: ability to estimate high-resolution social contact patterns from longitudinal surveys}

\author[1,*,+]{Shozen Dan}
\author[1]{Joshua Tegegne}
\author[1]{Yu Chen}
\author[1,2]{Zhi Ling}
\author[3]{Veronika K Jaeger}
\author[3]{André Karch}
\author[2,4]{Swapnil Mishra}
\author[1,*,+]{Oliver Ratmann} 
\author[]{on behalf of the Machine Learning \& Global Health network}
\affil[1]{Imperial College London, Department of Mathematics, United Kingdom}
\affil[2]{National University of Singapore, Saw Swee Hock School of Public Health, Singapore}
\affil[3]{University of Münster, Institute of Epidemiology and Social Medicine, Germany}
\affil[4]{National University of Singapore, Institute of Data Science, Singapore}

\affil[*]{shozen.dan21@imperial.ac.uk, oliver.ratmann@imperial.ac.uk}
\affil[+]{these authors contributed equally to this work}

\begin{abstract}
Social contact surveys are an important tool to assess infection risks within populations, and the effect of non-pharmaceutical interventions on social behaviour during disease outbreaks, epidemics, and pandemics. 
Numerous longitudinal social contact surveys were conducted during the COVID-19 era, however data analysis is plagued by reporting fatigue, a phenomenon whereby the average number of social contacts reported declines with the number of repeat participations and as participants' engagement decreases over time.
Using data from the German COVIMOD Study between April 2020 to December 2021, we demonstrate that reporting fatigue varied considerably by sociodemographic factors and was consistently strongest among parents reporting children contacts (parental proxy reporting), students, middle-aged individuals, those in full-time employment and those self-employed. 
We find further that, when using data from first-time participants as gold standard, statistical models incorporating a simple logistic function to control for reporting fatigue were associated with substantially improved estimation accuracy relative to models with no reporting fatigue adjustments, and that no cap on the number of repeat participations was required.
These results indicate that existing longitudinal contact survey data can be meaningfully interpreted under an easy-to-implement statistical approach adressing reporting fatigue confounding, and that longitudinal designs including repeat participants are a viable option for future social contact survey designs.
\end{abstract}
\begin{document}

\flushbottom
\maketitle
\thispagestyle{empty}

\section*{Introduction}
\input{1_introduction}

\section*{Results}
\input{2_results}

\section*{Discussion}
\input{4_discussion}

\section*{Methods}
\input{3_methods}

\bibliography{references}



\section*{Acknowledgements}
We thank the CoMix team and in particular Christopher Jarvis, Kevin Van Zandvoort, Amy Gimma, John Edmunds for sharing the CoMix questionnaire for the COVIMOD surveys, and for their cooperation; the team at IPSOS-Mori for their work and efforts on implementing the COVIMOD survey; the Imperial College Research Computing Service (https://doi.org/10.14469/ hpc/2232) for providing the computational resources
to perform this study; and Zulip for sponsoring team communications through the Zulip Cloud Standard chat app.

S.D. and Y.C. acknowledges funding from EPSRC Centre for Doctoral Training in Modern Statistics and Statistical Machine Learning at Imperial and Oxford (EP/S023151/1 to Prof. Axel Gandy); Z.L. is funded by the National Medical Research Council, Singapore Ministry of Health (PREPARE-S2-2024-002); 
O.R. acknowledges grant funding from the Bill \& Melinda Gates Foundation (OPP1175094), the Engineering and Physical Sciences Research Council (EP/X038440/1), the National Institute of Health (R01AI155080), and the Moderna Charitable Foundation; S.M. acknowledges support from the National Research Foundation, Singapore, under its NRF FELLOWSHIP (NRF-NRFF15-2023-0010).

COVIMOD was funded by intramural funds of the Institute of Epidemiology and Social Medicine, University of Münster, and of the Institute of Medical Epidemiology, Biometry and Informatics, Martin Luther University Halle-Wittenberg, as well as by funds provided by the Robert Koch Institute, Berlin, the Helmholtz-Gemeinschaft Deutscher Forschungszentren e.V. via the HZEpiAdHoc "The Helmholtz Epidemiologic Response against the COVID-19 Pandemic" project, the Saxonian COVID-19 Research Consortium SaxoCOV (co-financed with tax funds on the basis of the budget passed by the Saxon state parliament), the Federal Ministry of Education and Research (BMBF) as part of the Network University Medicine (NUM) via the egePan Unimed project (funding code: 01KX2021) and the Deutsche Forschungsgemeinschaft (DFG, German Research Foundation, project number 458526380).

The funders had no role in study design, data collection and analysis, decision to publish or preparation of the manuscript.

\input{supp_methods_1}
\input{supp_figure_1}
\input{supp_figure_2}
\input{supp_figure_3}

\section*{Author contributions statement}
S.D. and O.R. wrote the main manuscript text. S.D., J.T., Y.C., and Z.L. performed data analysis, analysed the results, and prepared figures. V.J. and A.K. provided the data and wrote the methods and discussion. S.M. wrote the introduction, methods, and discussion. All authors reviewed the manuscript.

\section*{Competing interests}
The authors declare no competing interests.

\section*{Data Availability}
Data can be provided promptly upon valid request to the University of Münster (contact: Veronika.Jaeger@ukmuenster.de) and will be distributed via a secure server. All the code required to reproduce the results can be found at: https://github.com/ShozenD/contact-survey-fatigue


\end{document}

%% file: 1_introduction.tex
Despite advances in medicine and public health, infectious respiratory diseases such as coronaviruses, diphteria, influenza, measles or tuberculosis continue to cause substantial global morbidity \cite{james_global_2018} and mortality \cite{roth_global_2018}. These pathogens are easily transmitted via close-range human-to-human contacts, which may be direct via physical contact or indirect through airborne droplets and aerosols \cite{brankston_transmission_2007, killingley_routes_2013}, and for this reason have high pandemic potential. Consequently, it is fundamental for pandemic preparedness to understand the structure and intensity of human social contacts -- across the full diversity of human populations globally\cite{mousa_social_2021}. Contact data are used to provide real-time estimates of key quantities such as epidemic reproduction numbers\cite{gimma_changes_2022}, to optimise vaccine allocation\cite{medlock_optimizing_2009}, to estimate transmission risk factors\cite{ferretti_quantifying_2020}, to provide nuanced assessments into the population groups that drive transmission\cite{monod_longitudinal_2024, chang_heterogeneity_2016, morris_concurrent_1995}, and identifying efficacious strategies to curb disease transmission \cite{liu_rapid_2021, coletti_comix_2020, gimma_changes_2022, tomori_individual_2021, wymant_epidemiological_2021}.

Protocols for collecting social contact data through surveys are open-source\cite{willem_socrates_2020}, and have been systematically implemented internationally during the COVID-19 pandemic\cite{wong_social_2023}. There are several data collection features that obscure the true structure and intensities of contacts, most notably that participants are reporting the age of their contacts in coarse age brackets; that contacts of individuals under the age of 18 years are frequently reported by caregivers; that the composition of the survey panels that comprise all participants is updated over time although individual-level survey weights are available; and that repeat participants who contribute to consecutive surveys tend to under-report their contacts, a phenomenon termed reporting fatigue~\cite{egleston_impact_2011, wong_social_2023}. 
Dan et al.~\cite{dan_estimating_2023} demonstrated it is possible to decipher contact intensities by 1-year age bands from coarse data due to constraints on how the actual contacts that occurred must add up, and also leverage data from adults on contacts with children to decipher the contacts of children. Post-stratification methods can also be used further to incorporate survey weights\cite{gimma_changes_2022, little_post-stratification_1993}. However, it remains unclear to what extent it is possible to adjust for the confounding effects of reporting fatigue after the data have been collected~\cite{dan_estimating_2023, loedy_longitudinal_2023, angela_sinickas_finding_2007, philip_cleave_survey_2022}, and consequently how future longitudinal contact surveys should best be implemented.

Here, we investigate contact data collected longitudinally over 33 survey waves from April 2020 to December 2021 in Germany through the COVIMOD Study to address the following questions: What are the individual determinants of reporting fatigue; is it possible to control for individual-level reporting fatigue in statistical models aimed at estimating the structure and intensity of social contacts; and is there an upper limit on repeat participation beyond which the error introduced through reporting fatigue becomes challenging to control for in statistical models? The COVIMOD surveys were conducted using the similar protocol as all countries participating in the European CoMix Study. For this reason, we expect our investigations will generally improve our ability to interpret social contact data already collected, our ability to design future longitudinal contact surveys that are robust to reporting fatigue effects, and improve accuracy in infectious disease incidence forecasting\cite{banholzer_comparison_2023}.

%% file: 2_results.tex
\begin{figure}[p!]
    \centering
    \includegraphics{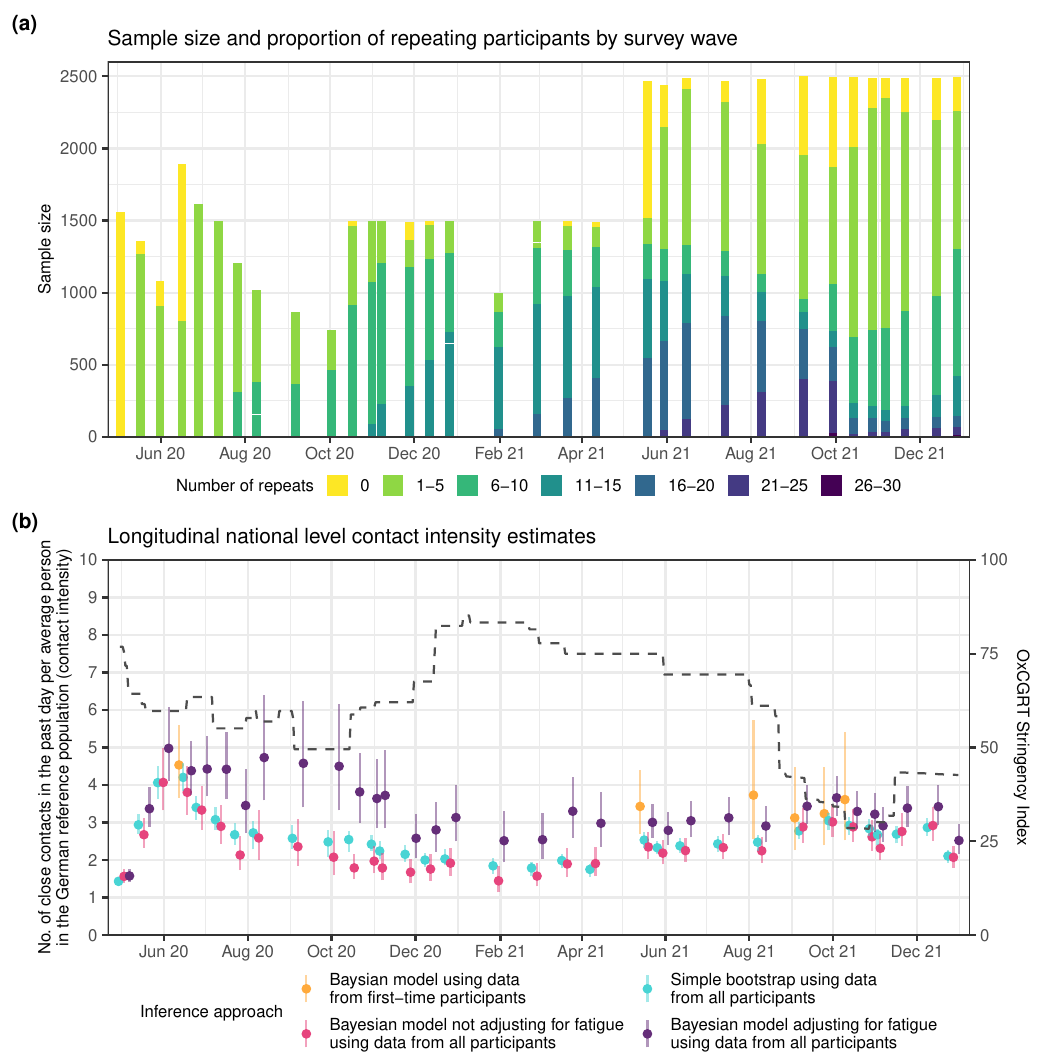}
    \caption{\textbf{Longitudinal contact intensity estimates during the COVID-19 pandemic in Germany.} (\textbf{a}) Longitudinal structure of 33 survey waves of the COVIMOD contact survey study, showing sample sizes for each survey wave on the y-axis. Colours indicate the number of repeat participations of each participant in previous survey waves. (\textbf{b}): Longitudinal, national-level contact intensity estimates (point: simple bootstrap mean or posterior median estimate, linerange: 95\% bootstrap confidence or 95\% credible intervals) are shown according to different estimation approaches: Bayesian model using data from first-time participants only, for waves with more than 300 first-time participants (orange); simple bootstrap using data from all participants and not accounting for reporting fatigue (blue), Bayesian model using data from all participants and not adjusting for reporting fatigue (pink); Bayesian model using data from all participants and adjusting for reporting fatigue (purple). The dashed line represents the OxCGRT Stringency Index with higher values indicating a higher degree of contact restrictions (min: 0, max: 100).}
    \label{fig:1}
\end{figure}

\subsection*{Determinants of reporting fatigue and pandemic contact intensities}
During the COVIMOD surveys, $7,851$ participants provided information on $141,928$ close social contacts they had in the previous day between April 30, 2020 and December 31, 2021.
Two survey waves of the COVIMOD study, wave 4 and 21, included a large number of first-time participants and repeating participants   (Figure~\ref{fig:1}a). This enabled us to identify measured individual-level features associated with variations in the number of close social contacts that one typical person in a population had per day (the contact intensity) during the pandemic, and following that to investigate the impact of reporting fatigue on these features among repeating participants that were surveyed at the same time point.
Specifically, wave 4 (June 11 to June 22, 2020) included 1,089 first-time participants and 801 repeating participants, and wave 21 (May 5 to May 24, 2021) included 947 first-time and 1,521 repeating participants.
We performed a Bayesian feature selection analysis with sparsity inducing priors~\cite{piironen_sparsity_2017} to identify features associated with changes in pandemic contact intensities, and features associated with a reduction in contact intensities in connection with repeat participation. We first analysed participants from wave 4, and then repeated the analysis independently on data from wave 21 to test the robustness in the features selected across different phases of the pandemic.
Well-known determinants of contact intensities are age~\cite{dan_estimating_2023, backer_contact_2024, wong_social_2023, gimma_changes_2022, tomori_individual_2021, loedy_longitudinal_2023}, sex~\cite{dan_estimating_2023, wong_social_2023, tomori_individual_2021}, and household size~\cite{wong_social_2023, tomori_individual_2021, gimma_changes_2022, loedy_longitudinal_2023}, and we always included these factors in our analysis. 
Potential further determinants that were previously investigated and also measured during the COVIMOD study included employment status~\cite{wong_social_2023, gimma_changes_2022}, presence of disease symptoms, day of week effects~\cite{tomori_individual_2021}, and urban-rural typology (rural, intermediate population density, or urban), and these factors were tested for association with variation in contact intensities in the variable selection analysis.

\begin{table}[!p]
\centering
\begin{tabular}{llllcccc} \toprule
 & & \multicolumn{2}{c}{\# Participants (\# repeating)} & & & & \\
 \cmidrule(lr){3-4}
Variable & Category & \makecell{Survey\\ Wave 4} & \makecell{Survey\\ Wave 21} & \makecell{Well- \\ known} & \makecell{Relevant to \\ contact \\ intensity$^{*1}$} & \makecell{Relevant to \\ reporting \\ fatigue$^{*2}$} \\
\midrule
\multirow{14}{*}{Age group} & Preschool (0-5)      & 61 (8)& 69 (33)& $\circ$ & {\it n.t.} & $\circ$ \\
          & Raised at home (0-5) & 17 (2)& 44 (15)& $\circ$ & {\it n.t.} & $\circ$ \\
          & 6-9                  & 34 (8)& 51 (28)& $\circ$ & {\it n.t.} & $\circ$ \\
          & 10-14                & 80 (25)& 112 (71)& $\circ$ & {\it n.t.} & $\circ$ \\
          & 15-19                & 103 (27)& 145 (72)& $\circ$ & {\it n.t.} & $\circ$ \\
          & 20-24                & 106 (16)& 151 (81)& $\circ$ & {\it n.t.} & $\times$ \\
          & 25-34                & 216 (54)& 277 (172)& $\circ$ & {\it n.t.} & $\times$ \\
          & 35-44                & 163 (68)& 233 (157)& $\circ$ & {\it n.t.} & $\times$ \\
          & 45-54                & 271 (139)& 353 (209)& $\circ$ & {\it n.t.} & $\circ$ \\
          & 55-64                & 318 (197)& 395 (250)& $\circ$ & {\it n.t.} & $\circ$ \\
          & 65-69                & 311 (161)& 382 (255)& $\circ$ & {\it n.t.} & $\circ$ \\
          & 70-74                & 118 (61)& 174 (128)& $\circ$ & {\it n.t.} & $\times$ \\
          & 75-79                & 48 (24)& 52 (34)& $\circ$ & {\it n.t.} & $\circ$ \\
          & 80-84                & 5 (3)& 9 (5)& $\circ$ & {\it n.t.} & $\circ$ \\
\midrule
\multirow{2}{*}{Sex} & Male                 & 965 (391)& 1288 (807)& $\circ$ & {\it n.t.} & $\circ$ \\
          & Female               & 886 (402)& 1159 (703)& $\circ$ & {\it n.t.} & $\times$ \\
\midrule
\multirow{5}{*}{\makecell[l]{Household\\size}}   & 1 person & 477 (334)& 728 (457)& $\circ$ & {\it n.t.} & $\circ$ \\
          & 2 person             & 435 (165)& 825 (465)& $\circ$ & {\it n.t.} & $\circ$ \\
          & 3 person             & 535 (199)& 497 (339)& $\circ$ & {\it n.t.} & $\times$ \\
          & 4 person             & 206 (53)& 276 (167)& $\circ$ & {\it n.t.} & $\circ$ \\
          & 5+ person            & 198 (42)& 121 (82)& $\circ$ & {\it n.t.} & $\times$ \\
\midrule
\multirow{11}{*}{Employment} & Full-time employed               & 508 (220)& 702 (423)& & $\circ$  & $\circ$ \\
           & Part-time employed               & 202 (96)& 261 (166)& & $\times$ & $\circ$ \\
           & Self-employed            & 89 (48)& 106 (69)& & $\circ$  & $\times$ \\
           & Student (15+)            & 193 (45)& 228 (130)& & $\circ$  & $\circ$ \\
           & Retired                  & 462 (248)& 615 (422)& & $\circ$  & $\times$ \\
           & Long-term sick           & 42 (21)& 59 (33)& & $\circ$  & $\circ$ \\
           & \makecell[l]{Unemployed\\(seeking)}     & 82 (32)& 91 (49)& & $\circ$  & $\circ$ \\
           & \makecell[l]{Unemployed\\(not seeking)} & 34 (10)& 50 (37)& & $\circ$  & $\circ$ \\
           & Stay-at-home parent         & 47 (30)& 59 (34)& & $\circ$  & $\circ$ \\ \midrule
\multirow{2}{*}{\makecell[l]{COVID-like \\ Symptoms}} & Yes & 468 (182)& 620 (315) & & $\times$ & {\it n.t.} \\
                     & No            & 1383 (611)& 1827 (1195)& & $\times$ & {\it n.t.} \\\midrule
\multirow{2}{*}{Day of week} & Weekday                & 1579 (569)& 2359 (1433)& & $\circ$  & {\it n.t.}\\
            & Weekend                & 272 (224)& 88 (77)& & $\circ$  & {\it n.t.} \\ \midrule
\multirow{3}{*}{Urban type} & Rural                   & 282 (119)& 376 (232)& & $\times$ & $\circ$ \\
           & Intermediate            & 705 (291)& 923 (564)& & $\circ$  & $\circ$  \\
           & Urban                   & 864 (383)& 1148 (714)& & $\times$ & $\times$ \\ \midrule
           & {\bf Total}             & 1851 (793)& 2447 (1510)&  &          & \\ 
           \bottomrule
\multicolumn{7}{l}{{\it n.t.}: Not tested, $\circ$: Selected, $\times$: Not selected} \\
\multicolumn{7}{l}{$^{*1}$ Relevant features are those which increase or reduce contact intensity by more than 5\% relative to the baseline.} \\
\multicolumn{7}{l}{$^{*2}$ Relevant features are those which reduce contact intensity by more than 5\% relative to the baseline.}
\end{tabular}
\caption{\textbf{Determinants of social contact intensities and reporting fatigue in Germany, June 11-22 2020 (wave 4) and May 5-24 2021 (wave 21).}}
\label{tab:varselect-variables}
\end{table}

\begin{figure}[p!]
    \centering
    \includegraphics[width=0.99\textwidth]{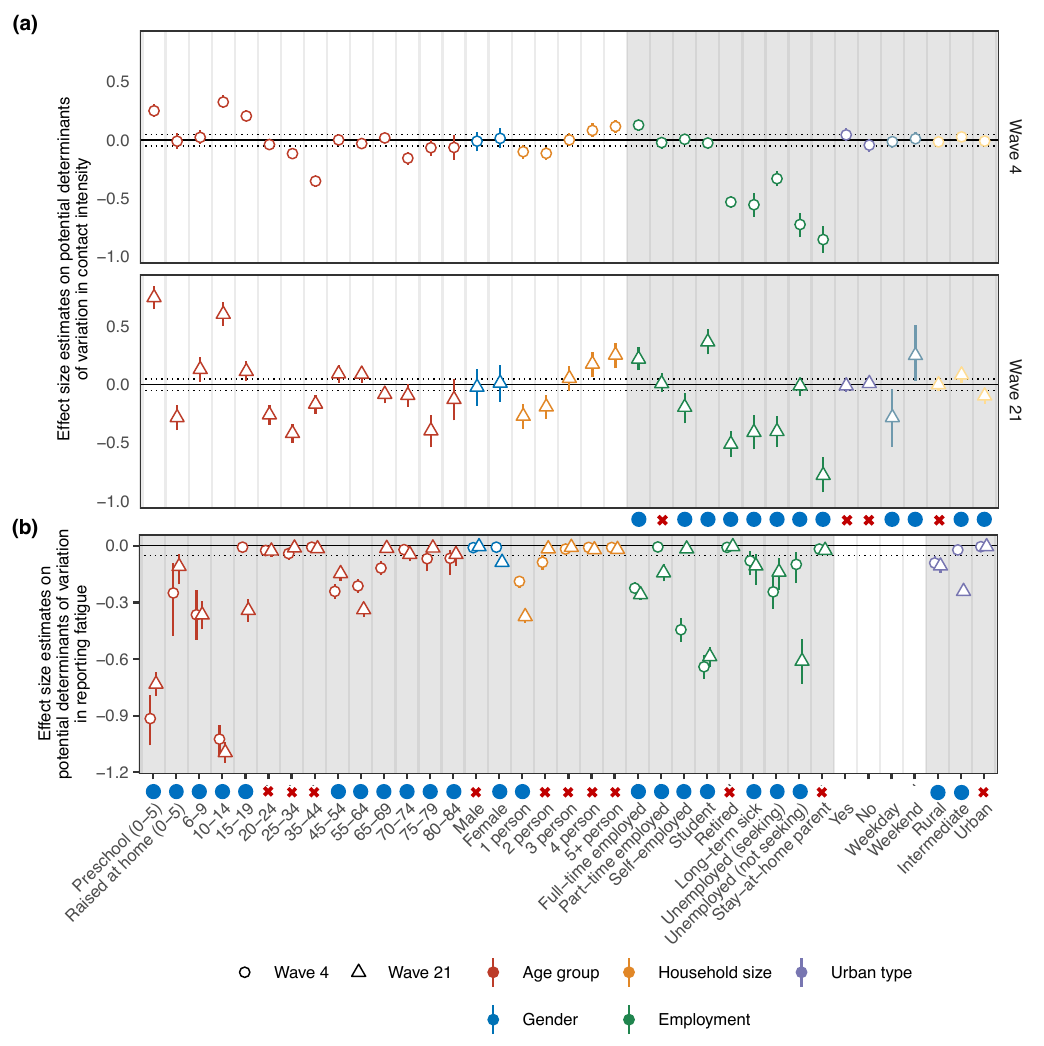}
    \caption{\textbf{Effect size estimates on determinants of variation in  contact intensity and reporting fatigue in the variable selection models.} (\textbf{a}) Posterior effect size estimates on contact intensity determinants in the variable selection model on survey data from survey waves 4 and 21 (circles and triangles: posterior median estimate in the Bayesian model using data from first-time participants, linerange: corresponding 50\% credible intervals) relative to the overall baseline contact intensity term in the model. Gray solid lines denote the average contact intensity, and gray dotted lines the variable selection cutoff threshold ($>\pm 5\%$ change). White background indicates features that were always included in the model, and grey background indicates features that were tested for inclusion through Bayesian variable selection. Blue solid dots denote selected variables and red crosses denote variables that were not identified to have $>\pm 5\%$ deviations in average contact intensities. (\textbf{b}) Posterior effect size estimates on reporting fatigue determinants in the variable selection model on survey data from survey waves 4 and 21 (points: posterior median estimate in the Bayesian model using data from repeating participants, linerange: corresponding 50\% credible intervals) relative to no effect. Other plot features are as in subfigure (a), except that the gray solid lines denotes no change in contact intensity as compared to first-time participants in the same subgroup, and the gray dotted line denotes the variable selection cutoff threshold ($> 5\%$ decrease).}
    \label{fig:2}
\end{figure}

Table~\ref{tab:varselect-variables} lists all features that we investigated, the total sample sizes for each respective subgroup in the two waves for first-time and repeating participants, the number of repeating participants (in brackets), and the identified determinants of contact intensities and reporting fatigue.
In our analyses, we stratified children aged 0-5 years by whether they were raised at home or attended pre-school due to the well-established influence of school attendance on child contact intensities~\cite{hens_estimating_2009}. 
Households of size 5 or more were grouped together, and comprised 10.7\% and 4.9\% of all households in the sample for survey wave 4 and wave 21, respectively.
In addition to age, sex, and households size, we found from the survey data of first-time participants that working full-time, being a student, living in areas of intermediate population density, and reporting on weekends were identified as features that increased sample-level average contact intensities by more than 5\% in the feature selection model (Figure~\ref{fig:2}a). Conversely, being self-employed, a stay-at-home parent, unemployed, retired, suffering from long-term sickness or being disabled, and reporting on weekday were identified as features that significantly decreased average pandemic contact intensities by more than 5\%. 
The features that were associated with >5\% deviations from average population-level contact intensities were qualitatively similar between the two waves (Figure~\ref{fig:2}a). Compared to wave 4, some of the identified features had a stronger influence on the number of contacts in wave 21, including being a student, the day of week in which the survey was answered, and population density. 

Next, we quantified the extent to which the reporting of contact intensities is impacted by reporting fatigue, defined as a reduction in contact intensities among individuals who participated in previous survey waves as compared to first-time participants in the same population subgroup. We extended the Bayesian regression model to include separate reporting fatigue effects for each of the previously selected contact intensity features, attached sparsity-inducing variable selection priors~\cite{piironen_sparsity_2017} to these, and then fitted the model to survey data from both first-time and repeating participants in wave 4, and finally to those in wave 21 to investigate the robustness of our findings (see Methods). We then retained reporting fatigue effects that were associated with a $>5\%$ reduction in posterior median contact intensities in repeating participants as compared to first-time participants in either survey wave 4 or wave 21. Table~\ref{tab:varselect-variables} reports the set of selected features, and Figure~\ref{fig:2}b shows the estimated percent reduction in contact intensity in repeating participants compared to first-time participants. Almost all contact intensity factors were associated with reporting fatigue, although to different extents. Again, the features that were associated with >5\% decreases from the contact intensities among first-time participants in the same population sub-group differed only slightly between the two waves (Figure~\ref{fig:2}b).

\begin{figure}[!ht]
    \centering
    \includegraphics[width=\textwidth]{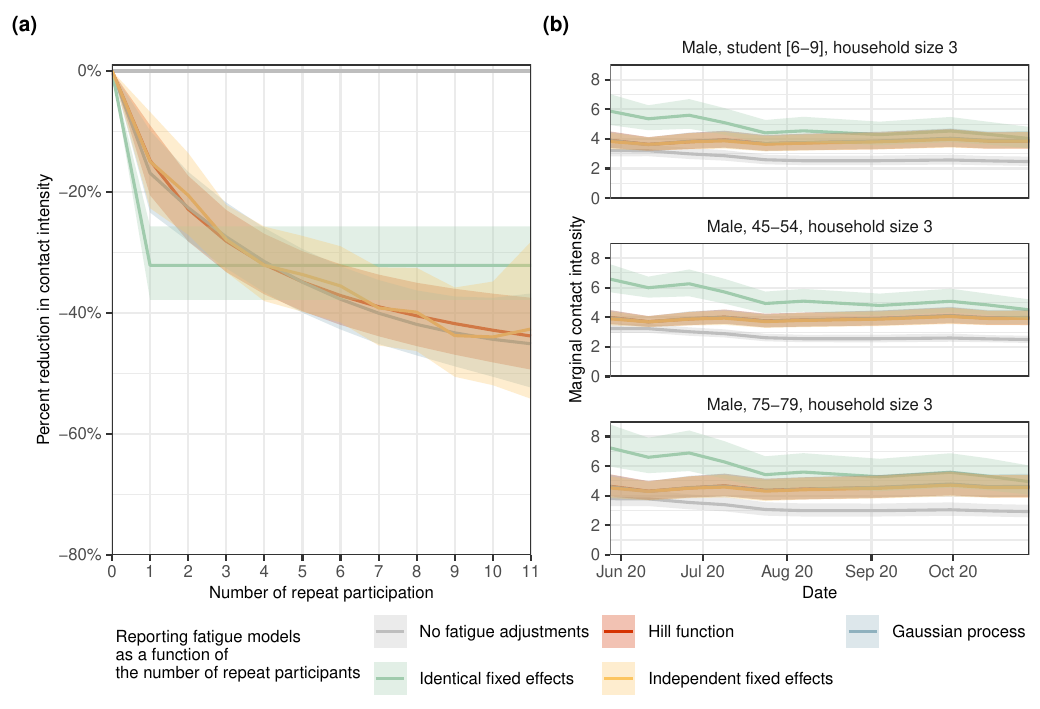}
    \caption{\textbf{Dynamics in the severity of reporting fatigue with increasing repeat participation in the COVIMOD Study.} (\textbf{a}) Estimated percent reduction in contact intensity as a function of the number of repeat participation for no reporting fatigue adjustments (gray), the Hill model (red), the Gaussian process model (blue), the identical fixed effects model (green), and the independent effects model (yellow) (line: posterior median estimate, ribbon: 95\% credible interval). (\textbf{b}) Estimated longitudinal contact intensities for men aged 6-9, 45-54, and 75-79 years living in a 3-person household for no reporting fatigue adjustments (gray), the Hill model (red), the Gaussian Process model (blue), the identical fixed effects model (green), and the independent effects model (yellow)}
    \label{fig:3}
\end{figure}

\subsection*{Impact of the number of repeat participations on reporting fatigue severity}
We next examined how reporting fatigue changes as a function of the number of repeat participations, using data from first-time and repeat participants from waves 3-12 (June to October, 2020), for which $n=1,287$ individuals had not participated previously, $n=8,057$ for 1-5 times, $n=3,397$ for 6-10 times, and $n=90$ for 11 times. Figure~\ref{fig:3}a shows in yellow, the percent reductions in reported contact intensities under a dynamic model that separately estimates distinct fixed fatigue effects for participants that repeated $r=0,1,\dotsc, 11$ times. It is clear that reporting fatigue resulted in under-reporting effects that were increasingly detrimental with repeat participation.
To develop a parsimonious model of reporting fatigue dynamics, we assessed three alternative modeling approaches: a identical fixed effects model~\cite{loedy_longitudinal_2023}, a non-parametric Gaussian Process (GP) model, and a parametric Hill model frequently used in bio-chemistry applications~\cite{prinz_hill_2009}. While the identical fixed effects model over-simplified reporting fatigue dynamics, the remaining two approaches gave similar results (Figure~\ref{fig:3}a) that also match results from the independent fixed effects model, indicating that fatigue dynamics can be well-approximated with Hill functions and it is not necessary to resort to more computationally intensive non-parametric techniques. The posterior median and 95\% CI for the parameters of the Hill function were $\gamma$: 0.88 [0.56, 1.74], $\zeta$: -1.55 [-2.26, -0.94], and $\eta$: 0.94 [0.59, 1.47], indicating that the maximum percentage reduction in the number of reported contacts due to repeating participation is approximately $100 \times (e^{-\gamma} - 1) = -58.5\%\ [-82.4\%, -42.9\%]$.
Figure~\ref{fig:3}b illustrates the impact of accounting for reporting fatigue dynamics on longitudinal contact intensity estimates in three population subgroups, men of age 6-9, 45-54, and 75-79 years living in a 3-person household. With the exception of the identical fixed effects model, all models with reporting fatigue adjustments suggest that contact intensities remained steady through waves 3 to 12 (June to October, 2020), in contrast to unadjusted estimates that exhibit notable decreases as the proportion of repeating participants increases (Figure ~\ref{fig:1}). 

\begin{figure}[p!]
    \centering
    \includegraphics[width=\textwidth]{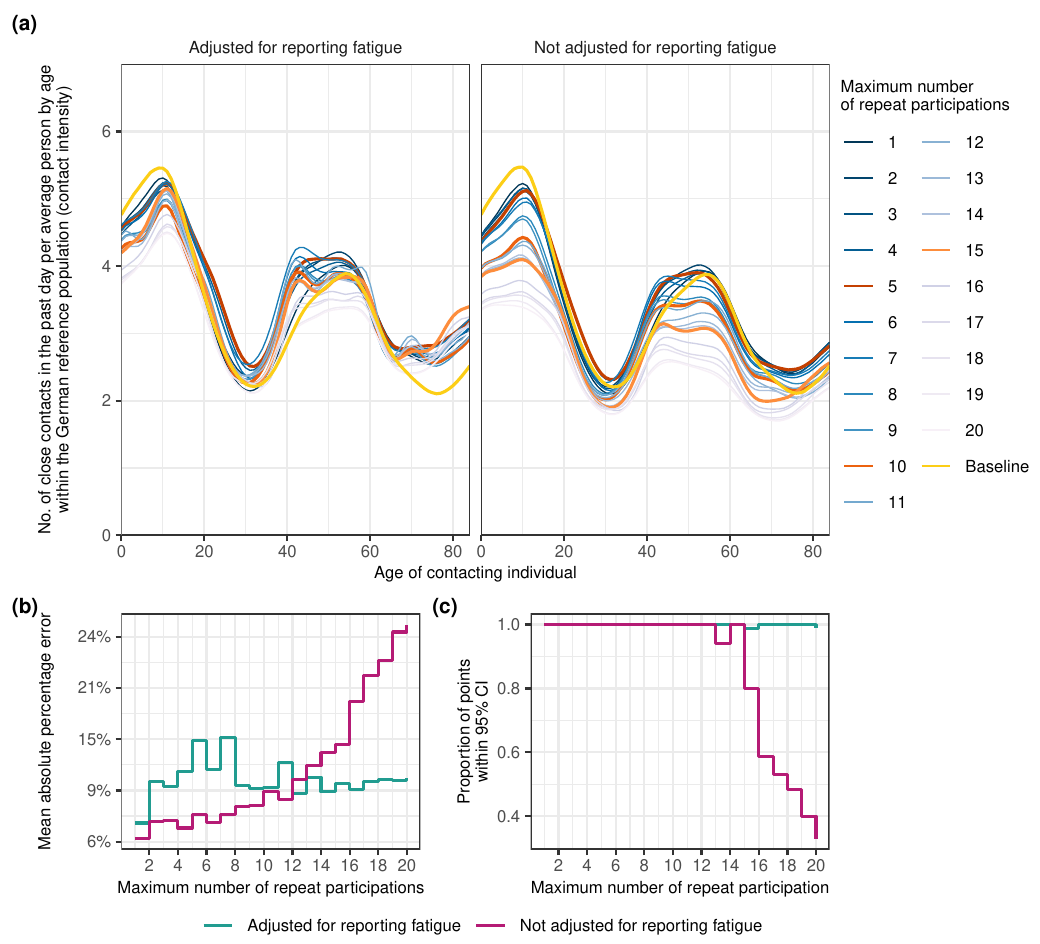}
    \caption{\textbf{Accuracy of the reporting fatigue Hill model in de-biasing age-specific contact intensity estimates from wave 21 of the COVIMOD Study.} (\textbf{a}) Posterior median contact intensity estimates for ages 0 to 84 during the COVIMOD survey wave 21, obtained from the reporting fatigue Hill model that adjusts for population subgroup-specific reporting fatigue effects via separate Hill functions (left) and the same model without reporting fatigue adjustments (right). In both panels, the gold curve represent posterior median estimates obtained from first-time participants. Coloured curves represent the posterior median estimates obtained from including participants with increasingly larger maximum number of repeats participations. (\textbf{b}) Mean absolute percentage error between the baseline (gold curve) and estimates obtained from the reporting fatigue Hill model on data including participants with more repeats (turquoise) and from the same model without reporting fatigue adjustments(purple). (\textbf{c}) Proportion of posterior median age-specific contact intensity estimates from data of first-time participants that fell within the 95\% credible intervals of the posterior age-specific estimates of the reporting fatigue Hill model on data including participants with more repeats.}
    \label{fig:4}
\end{figure}

\subsection*{Accuracy in de-biasing contact intensity estimates from reporting fatigue effects}
Given the substantial detrimental impacts of reporting fatigue on longitudinal contact intensity estimates, we next assessed the possibility of de-biasing contact intensity estimates by  utilising the Hill function approach from the previous Section. In the experiment, we aimed at de-biasing age-specific contact intensities as non-pharmaceutical interventions often target specific age groups~\cite{hens_estimating_2009}, and compared the accuracy in age-specific contact intensities inferred from data from participants with increasingly many repeat participations to the age-specific contact intensities inferred from first-time participants. We used data from survey wave 21 for which the sample consists of a large proportion of first-time participants and participants with a large number of repeats. We extended our Bayesian model to specify age-specific contact intensities with a Gaussian process on the age of survey participant, and assigned independent Hill functions to every feature selected in the previous analysis (Methods). The model was applied first to data with first-time participants only to serve as a baseline which excludes the confounding effects of reporting fatigue. Data on contacts in wave 21 from individuals who participated in previous survey rounds were then incrementally added back into the sample to understand the ability of the reporting fatigue Hill model to de-bias contact intensity estimates. Figure~\ref{fig:4}a compares the posterior median age-specific contact intensity estimates from first-time participants (yellow) to those obtained by incrementally including groups of participants with more repeat participations, in both a model that adjusts for reporting fatigue (left) and a model without any reporting fatigue adjustments (right). The mean absolute percentage error (MAPE) between the age-specific contact intensity estimates from first-time participants and those from data including participants with increasingly many repeat participations is shown in Figure~\ref{fig:4}b, and reveals a marked difference in performance between the reporting fatigue Hill model and the model without reporting fatigue adjustments. The MAPE from the unadjusted model increased steadily with the maximum number of repeat participations and reaches approximately 24\% with the inclusion of all participants, whereas estimates from the reporting fatigue Hill model consistently had a MAPE around 9\%-15\%. Figure~\ref{fig:4}c examines the proportion of the age-specific contact intensity estimates from first-time participants that fell within the 95\% posterior credible intervals of the age-specific estimates obtained from data including participants with increasingly many repeat participations. In the models without adjustment for reporting fatigue, coverage decreased noticeably after incorporating data from participants with 15 repeats. In contrast, the reporting fatigue Hill model maintained $\approx$100\% coverage regardless of the maximum number of repeat participations. Past these global assessment measures, Figure~\ref{fig:4}a indicates estimates were most sensitive to reporting fatigue in age ranges 0-15 years and 40-59 years which are highly relevant in disease transmission~\cite{hens_estimating_2009}. In models without reporting fatigue adjustments, capping data to individuals with up to 5, 10, and 15 repeat participations resulted in a MAPE of 7.0\%, 12.8\%,  18.0\% leaving out 52.3\%, 16.8\%, and 5.8\% of responses, respectively. In contrast, in the Hill model with reporting fatigue adjustments, capping at the same thresholds resulted in a MAPE of 9.6\%, 7.4\%, and 9.9\% in these age groups.

\subsection*{Longitudinal and subgroup-specific contact intensity estimates during the pandemic}
\begin{figure}[ht!]
    \centering
    \includegraphics[width=\linewidth]{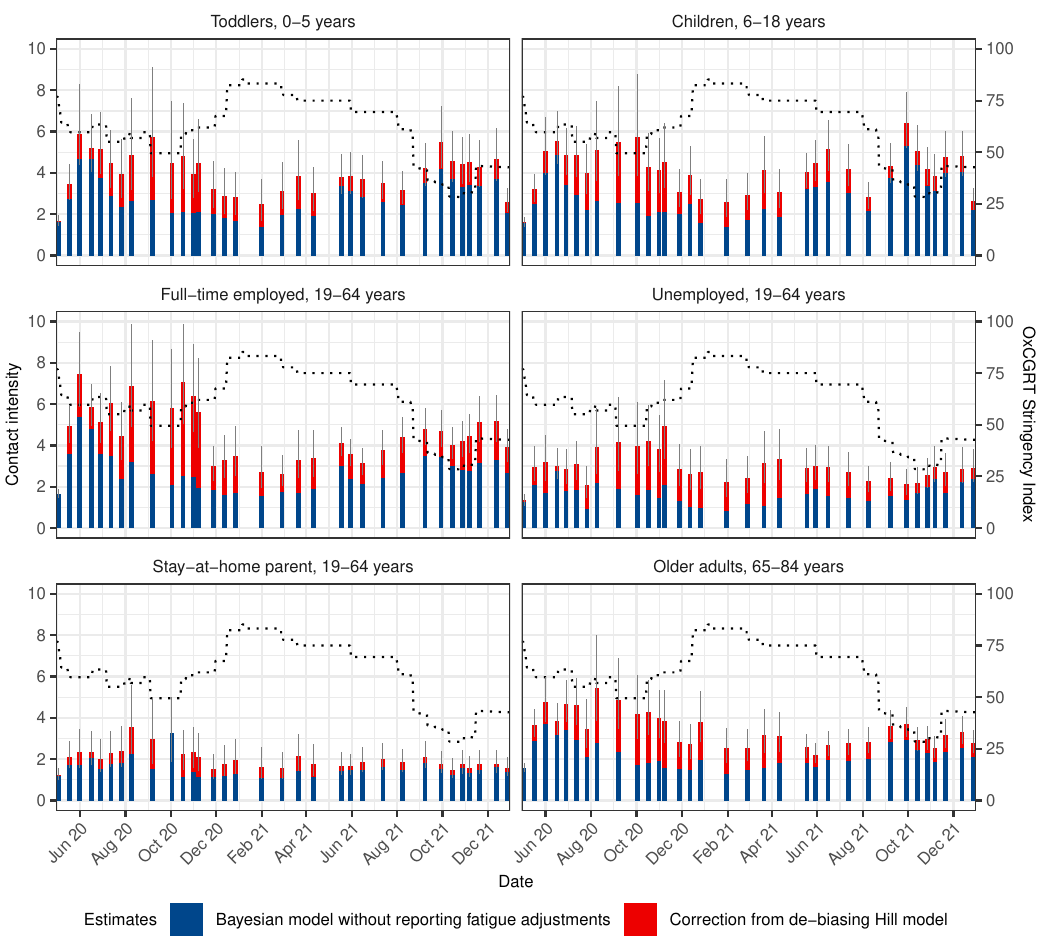}
    \caption{\textbf{De-biased longitudinal contact intensity estimates during the COVID-19 pandemic in Germany.} Posterior median contact intensity estimate corrections for eight different population subgroups from the de-biasing Hill model (turquoise) on top of posterior median contact intensity estimates derived without reporting fatigue adjustments (orange). 95\% credible intervals of the debiased contact intensity estimates from the Hill model are shown as lineranges. The purple dashed lines denote the OxCGRT stringency index~\cite{hale_global_2021}. Percent relative corrections of estimates from the Hill model against the Bayesian model without reporting fatigue adjustments are given in Supplemental Figure 2.}
    \label{fig:5}
\end{figure}

With this accuracy assessment in place, we applied the de-biasing Hill model to correct for reporting fatigue in the empirical survey data, and obtained under-reporting adjusted, national-level contact intensity estimates across Germany from April 2020 to December 2021 (survey waves 1 to 33).
Specifically, we used the de-biasing Hill model with all selected determinants of contact intensities and reporting fatigue as described in Table~\ref{tab:varselect-variables}, group-specific Hill functions and hyperparameters, and non-parametric, smooth age-specific Gaussian process offsets as shown in Figure~\ref{fig:4}a  to data from all survey waves of the COVIMOD study and from all first-time and repeating participants (Methods).
Figure~\ref{fig:1}b shows the posterior median population-level average contact intensities and 95\% credible intervals for all survey waves estimated by different inference approaches and models. In most survey waves (except survey wave 25, discussed below), the de-biased contact intensity estimates were very close to those obtained with the same model on data from first-time participants (purple), and those obtained with the simple bootstrap algorithm on data from first-time participants (Supplementary Fig. 1). 
%
In contrast, contact intensity estimates obtained with the same model but without reporting fatigue Hill function adjustments on data from all participants (yellow), or contact intensity estimates obtained with the \texttt{socialmixr}~\cite{funk_socialmixr_2024} methodology on data from all participants were clearly subject to notable under-reporting. We further compared estimates from the de-biasing Hill model and the same model without reporting fatigue adjustments for several population sub-groups. Reconsidering our findings in Figure~\ref{fig:2}b, we expected that contact intensity estimates for particular sub-populations are particularly strongly affected by reporting fatigue under-reporting, including preschool toddlers aged 0-5 years, children ages 6-18 years, students, or self-employed adults. We found that the de-biasing Hill model substantiated these expectations (Figure~\ref{fig:5}, Supplemental Figure 2), and typically provided non-trivial corrections to longitudinal contact intensity estimates that resulted primarily from the non-trivial structure of repeat participants in the cohort.

%% file: 4_discussion.tex
In this work, we tackled three pivotal questions aimed at enhancing our ability to understand temporal changes in social contact in a pandemic from longitudinally collected surveys: First, what are the primary individual characteristics associated with reporting fatigue; second, is it possible to de-bias contact intensity estimates from under-reporting effects due to reporting fatigue; and third, what is the impact on contact estimates throughout the COVID-19 pandemic when we adjust for reporting fatigue. Our findings indicate that reporting fatigue is particularly relevant in children, middle age individuals, and the working population (Figure~\ref{fig:2}, ~\ref{fig:4}). We identify the functional form of reporting fatigue and introduce the use of Hill functions as a viable modelling method to mitigate its detrimental effects (Figure~\ref{fig:3}). We show that the effects of reporting fatigue differs across different population subgroups and also changes dynamically over time (Figure~\ref{fig:5}). Our model's adjustments for reporting fatigue suggest that social contact intensities were higher than those estimated by non-adjusting methods, and that these adjustments brings estimates closer to those estimated using data from first-time participants alone (Figure~\ref{fig:1}, Figure~\ref{fig:4}).

Recruiting participants multiple times in longitudinal web-based social contact surveys is necessary as opposed to enlisting a new panel for each wave. In particular, the limited number of members within any given online panel necessitates repeated sampling, as it is only a matter of time before all eligible and willing participants have been recruited. This limitation is particularly challenging when striving for a sample that is approximately representative of the population. While panels from various market research companies could theoretically be employed to circumvent this issue, in practice, it requires each company to independently recruit participants and administer surveys within their panels, leading to substantial organisational, financial, and temporal costs. Allowing participants to be enrolled multiple times not only mitigates these costs but also facilitates the attainment of a sample that closely mirrors the population age and sex structure. Moreover, repeated sampling enables the examination of intra-personal variability in contact behavior over time, an area largely unexplored in pre-pandemic studies~\cite{mossong_social_2008}. Despite these benefits, allowing for multiple participation presents significant challenges in the form of reporting fatigue. Multiple studies across different countries have shown that on average, the relative number of contacts reported by individuals decrease with the number of participations in the survey~\cite{wong_social_2023, dan_estimating_2023, loedy_longitudinal_2023}. This limits our ability to accurately understand contact dynamics which in turn limits our ability to accurately assess infection risk and estimate reproduction numbers~\cite{goeyvaerts_estimating_2010, hamilton_examining_2024, stanke_estimating_2023}, which is crucial for targeted non-pharmaceutical interventions, vaccinations strategies, and preparations for future pandemics.

Our results suggest that parents reporting children contacts (parental proxy reporting), students (aged 19+), middle-aged individuals (aged 40-59, those in full-time employment and those self-employed were the subgroups most effected by reporting fatigue. We postulate that the relative reduction in reported contacts among children is primarily due to the reporting fatigue experienced by parents, who report on behalf of their children. Moreover, parents face notable challenges in reporting their children's social interaction at school and other educational settings, where direct parental supervision is uncommon. This significantly hinders parents' ability to accurately account for all of their children's contacts, even under the best circumstances. Students and middle-aged working individuals may be the subgroups most affected by recall bias as they may find it hard and/or tiresome to recall and report a large number of contacts at school/work place and other non-household settings. Designers of future social contact studies may benefit from these observations and take informed design decisions to mitigate bias in these specific subgroups.

For large-scale pandemic-era social contact studies, which must be conducted online for efficiency in scale and timing, implementing thorough quality control may not always be feasible, and no amount of meticulous design may prove sufficient to eliminate reporting fatigue bias. For data that has already been collected, modelling is necessary to correct for the biases discussed above. Our longitudinal models applied data from wave 3 to 12 (Figure~\ref{fig:3}) and empirical results from past studies~\cite{wong_social_2023} suggests that fatigue effects should be modelled as a monotonically decreasing non-linear function of the number of repeat participation. While numerous candidates exist, we propose using the Hill function for its intuitive interpretation as a dose-response model, where administering the survey is considered a dose and fatigue is the response~\cite{prinz_hill_2009}. Despite imposing strong assumptions on the functional form of fatigue effects, the Hill function offers computational advantages over non-parametric alternatives like Gaussian processes, as it requires estimating only three parameters. An application of the Hill function approach on data from wave 21 shows that it can effectively mitigate the effects of reporting fatigue. This allows us to make use of a larger portion of the collected data with less fear of obtaining significantly under-estimated values.

The findings of this study should be considered in the context of several limitations. First, with regards to the collected data and despite quota sampling, the final COVIMOD samples were not fully representative of the German population. Middle-aged individuals with children were underrepresented, as many surveys were completed by parents on behalf of their children. Moreover, COVIMOD was conducted as an online survey, with participants recruited from an online marked research panel through email, which may introduce bias as those with internet access who take part in health-related surveys might exhibit a higher compliance with social distancing measures. The retrospective nature of the reporting may also reduce the accuracy of reports as participants face challenges in recalling past events. Further details on other potential sources of bias in COVIMOD and strategies employed to address them is available elsewhere\cite{tomori_individual_2021}. Second, with regards to our methodological approach, accuracy estimates were obtained using data on first-time participants as gold-standard for reference. However, these participants may not always be representative of the sample or population. Despite this limitation, we argue that using first-time participants' data is the most viable option in the absence of a definitive method to establish the absolute truth. Additionally, we applied Hill function-based adjustments for each variable identified during the feature selection stage, allowing us to explore differences in reporting fatigue across various population subgroups (Figure~\ref{fig:5}). However, as indicated in Figure~\ref{fig:1}, the variables associated with reporting fatigue may vary across waves and  are additionally potentially heterogeneous across individuals in terms of non-measured factors. Future research is needed to investigate the extent of individual-level reporting fatigue effects beyond the co-variates that we considered. For instance, we can treat data from first-time and repeating participants as separate datasets and apply semi-modular inference (SMI)~\cite{carmona_semi-modular_2020} to infer social contact estimates. This approach can mitigate contamination from repeating participants' but requires careful calibration of the balance between information from both groups. Third, in terms of our central findings, we only had sufficient numbers of first-time and repeating participants in two survey waves in order to assess reporting fatigue effects and the accuracy of the de-biasing Hill model through our semi-modular study design. Future work should include further investigations on other longitudinal survey data such as that from the CoMIX Study. For waves with a small number of repeats, modeling fatigue using independent fixed effect terms may still be a viable and practical approach. However, based on our findings, we caution against capping the number of participant repeats at a certain threshold (e.g., 5 or 10) and proceeding to estimate contact intensity without model-based adjustments, as previous studies~\cite{dan_estimating_2023, loedy_longitudinal_2023} and our analysis (Figure~\ref{fig:1}, Figure S1 demonstrate that excluding participants with a small number of repeats can lead to significant underestimation of contact estimates.

In conclusion, this study identifies key individual-level features influencing overall contact intensity, and those associated with reporting fatigue. We find simple-to-implement Hill functions can accurately control for reporting fatigue effects that originate commonly in contemporary longitudinal social contact survey designs. While this methodology may not completely eliminate the bias induced by reporting fatigue, it enables us to use the entire dataset with reduced concern about underestimating contact intensity. Our findings highlight that existing longitudinal contact survey data can be meaningfully interpreted despite reporting fatigue effects, and that longitudinal designs including repeat participants are a viable option for future social contact survey designs. By refining the accuracy of contact estimates, model-based statistical approaches can better inform public health strategies and interventions at high resolution beyond age and gender, and thereby contribute to more effective management and mitigation of contact-related infectious disease outbreaks in the future.

%% file: 3_methods.tex
\subsection*{The COVIMOD study}
The COVIMOD study is a repeated survey from April 2020 to December 2021 comprising 33 waves. Participants were recruited through email invitations sent to existing panel member of the online market research platform IPSOS i-Say \cite{noauthor_ipsos_nodate}. To ensure the sample's broad representativeness of the German population, quota sampling was conducted based on age, sex, and region. A subgroup of adult participants living with children under the age of 18 were selected to answer the survey on behalf of their children. Similar to the CoMix study\cite{gimma_changes_2022, coletti_comix_2020}, individuals were invited to participate in multiple waves to track changes in behaviour and attitude toward COVID-19 across time. When the participant size did not meet the sampling quota due to dropout, new participants were recruited into the study.

The COVIMOD questionnaire was based on the questionnaire of the CoMix study\cite{gimma_changes_2022, coletti_comix_2020}, and includes questions on demographics, the presence of a household member belonging to a high-risk group, attitudes towards COVID-19 as well as related government measures~\cite{verelst_socrates-comix_2021, tomori_individual_2021}. Participants were asked to provide information about their social contacts between 5 a.m. the previous day to 5 a.m. the day of answering the survey. A contact was defined as either a skin-to-skin contact (e.g., kiss, handshake, hug), or the exchange of three or more works in the presence of another person. Participants reported information on the sex and age band of each contact, their relationship, the contact setting (e.g., home, school, workplace, place of entertainment), and whether the contact was a household member.

For survey waves 1 and 2, participants were asked to report information on each contact separately. However, from wave 3 participants were provided with the option to report the total number of contacts in the event that they could not list all of them separately. Additionally, some participants could not recall or preferred not to answer the age or sex information of some of the individual contacts. We treat these three types of entries with missing age or sex equally. A copy of the COVIMOD questionnaire may be found in Additional file 1 of Tomori et al. 2021 \cite{tomori_individual_2021}. COVIMOD was approved by the ethics committee of the Medical Board Westfalen-Lippe and the University of Münster, reference number 2020-473-fs. Written consent was obtained from all participants.

\subsection*{Other datasets}
To better contextualise the changes in social contact patterns in relation to non-pharmaceutical interventions, we acquired data from the Oxford Covid-19 Government Response Tracker (OxCGRT) which collected information on policy measures that tackle COVID-19 through 2020 to 2021~\cite{hale_global_2021}. Specifically, we extract the OxCGRT compact dataset for Germany and use the stringency index which measures containment, closure policies, and public information campaigns. The urban-rural typologies of where participants lived were determined by matching residence information to nomenclature of territorial units for statistics (NUTS) level 3 regions and using the classification defined by Eurostat. Specifically, predominantly urban regions are where at least 80\% of the population live in urban clusters (a urban-rural typology of at least 300 inhabitants per km² and a minimum population of at least 5000 inhabitants); intermediate regions are defined as regions where between 50\% but to 80\% of the population live in urban clusters; and predominantly rural regions are defined as regions where least 50\% of the population live in rural areas (all areas outside urban clusters)~\cite{noauthor_territorial_2024}. NUTS data for 2021 were downloaded from the Eurostat website~\cite{noauthor_territorial_nodate}. To construct poststratification weights we obtained population size estimates by age and sex for 2019 (Table code: 12411-0041) and household size estimates for 2019 (Table code: 12211-9020) from the GENESIS-online database~\cite{noauthor_genesis-online_2024}.

\subsection*{Software}
All analysis in this work was performed using R version 4.4.1~\cite{r_core_team_r_2024}.
Bayesian inference was performed using the \textit{Stan}\cite{stan_development_team_stan_2022} probabilistic programming language through the \textit{cmdstanr}\cite{jonah_gabry_cmdstanr_2023} package version 2.34.1 as front-end. A number of analyses were performed on high performance computing clusters maintained by the research computing service at Imperial College to reduce experiment time but all analysis can be run on modern laptops. The data and code to replicate the results in this study can be found at: https://github.com/ShozenD/contact-survey-fatigue.

\subsection*{Data processing}
In all analyses, we omitted data for participants who did not disclose their age or sex, information which is required in all of our models. Adhering to ethical standards, the age of participants under 18 years (children) were reported in distinct age groups, namely, 0–4, 5–9, 10–14, and 15–18 years. To ascertain detailed age data for those under 18, we imputed their age by selecting from a discrete uniform distribution framed by the minimum and maximum ages of the participant's age group. The total number of contacts reported by individuals were truncated to 30 to mitigate the effects of extreme outliers on statistical estimates.

\subsection*{Features associated with pandemic social contacts}
The number of contacts by an individual within a given time-frame can influenced by a variety of individual-level factors including age~\cite{dan_estimating_2023, backer_contact_2024, wong_social_2023, gimma_changes_2022, tomori_individual_2021, loedy_longitudinal_2023}, sex~\cite{dan_estimating_2023, wong_social_2023, tomori_individual_2021}, household size~\cite{wong_social_2023, tomori_individual_2021, gimma_changes_2022, loedy_longitudinal_2023}, employment status~\cite{wong_social_2023, gimma_changes_2022}, and day of week effects~\cite{tomori_individual_2021}. We perform a feature selection analysis with a focus on employment status (full-time employed, part-time employed, stay-at-home parent, long-term sick, retired, self-employed, student (above age 15), unemployed and seeking job, unemployed and not seeking job), day of week effects (weekday, weekend), presence of COVID-like symptoms (yes, no), and urban-rural typology (rural, intermediate, urban) while controlling for the effects of age, sex, and household size (Table~\ref{tab:varselect-variables}). To prevent reporting fatigue effects from contaminating the results, only records from first-time participants were used in the analysis. 

Let $Y^0_i$ for $i = 1,2,\ldots, n_0$ denote the number of contacts by a first-time participant. The Bayesian regression model has the form,
\begin{equation}\label{eq:varselect-first-time}
Y^0_i \sim \text{Poisson}(\lambda_i), \quad
\log(\lambda_i) = \beta_0 + \bm{u}^\top_i \alpha + \bm{v}^\top_i \beta, \quad
\quad \text{for } i = 1,2,\ldots,n_0.
\end{equation}
$\bm{u}_i$ is a vector of size $21 = 14 + 2 + 5$ containing the well-established features in the social contact literature which we do not test {\it i.e.} age, sex, and household size. $\bm{v}_i$ is a vector of size $16 = 9 + 2 + 2 + 3$ containing the features of interest: employment status, COVID-like symptoms, day of week, and urban-rural typology. We assigned generic un-informative priors to $\beta_0$ and $\alpha = (\alpha_1, \alpha_2, \ldots, \alpha_J)^\top$, 
\begin{align*}
    \beta_0 &\sim \text{Normal}(0, 100) \\
    \alpha_j &\sim \text{Normal}(0, \sigma^2_\alpha) \quad j = 1,2,\ldots,J \\
    \sigma_\alpha &\sim \text{half-Cauchy}^+(0, 1),
\end{align*}
whereas $\beta = (\beta_1, \beta_2, \ldots, \beta_{19})^\top$ is assigned a regularised horseshoe prior (RHS)~\cite{piironen_sparsity_2017} which pulls the marginal posterior of individual parameters with weak associations with the outcome toward 0 but allows those with strong associations to escape the shrinkage. The RHS prior is defined as,
\begin{subequations}
\begin{align}
\beta_k | \zeta_k, \varepsilon, c &\sim \text{Normal}(0, \varepsilon^2\tilde{\zeta}_k^2), \quad \tilde{\zeta}_k^2 = \frac{c^2\zeta_k^2}{c^2 + \varepsilon^2\zeta^2_k} \\
\zeta_k &\sim \text{student-}t^+_{\nu_1}(0,1), \quad k = 1,\ldots,K \\
c^2 &\sim \text{Gamma}(\nu_2, \nu_2 s^2/2) \\
\varepsilon &\sim \text{student-}t^+_{\nu_3}(0, \varepsilon_0),
\end{align}
\end{subequations}
where the scaling parameter $\varepsilon_0$ for $\varepsilon$ is expressed as 
\begin{equation*}
    \varepsilon_0 = \frac{p_0}{P-p_0} \frac{1}{n},
\end{equation*}
with $P$ the number of covariates in the model and $p_0 \in \{0, \dotsc, K\}$ specified to reflect prior knowledge on the effective number of non-zero coefficients~\cite{piironen_sparsity_2017}. For succinctness, we denote this prior as
\begin{equation*}
    \beta_k \sim \text{RHS}_{\nu_1,\nu_2,\nu_3}(s^2, p_0).
\end{equation*}
We set the hyperparameters of the RHS prior to $\nu_1 = 3, \nu_2 = 2, \nu_3 = 4, s^2 = 2$ to facilitate sampling in {\it Stan}~\cite{piironen_sparsity_2017}, and $p_0 = K/2$ to obtain a weakly informative prior on the number of selected variables. We also performed a sensitivity analysis where $p0$ was set to 1 (strong shrinkage) or $K - 1$ (weak shrinkage), however both settings did not change the results. Following inference, feature selection is performed given a relevance threshold. We set an optimistic threshold where if the value of the median of marginal posterior of the parameter is outside the range $(-0.0513, 0.0488)$ which corresponds to a $\pm 5\%$ change from the baseline parameter $\beta_0$. Additionally, features associated with social contacts may evolve over time with the introduction or the conclusion of non-pharmaceutical interventions, vaccination, public information campaigns, and other external factors. Therefore, in this and the following analysis, we applied the model independently to waves 4 and 21 which comprised of a balanced number of first-time and repeating participants (Figure~\ref{fig:1}), allowing us to compare the two groups at the same point in time. The final set of selected features was the union of the features selected in each wave.

\subsection*{Features associated with reductions in reported contacts in association to repeat participation}
The individual features associated with reporting fatigue, while identified as a issue in several studies~\cite{wong_social_2023, loedy_longitudinal_2023} have not been examined in detail. We investigate associations between reporting fatigue and schooling/age (14 categories), sex (2), household size (5), employment status (9), and urban-rural typology (3) (Table\ref{tab:varselect-variables}).

Let $Y^1_i$ for $i = 1,2,\ldots,n_1$ denote the number of contacts reported by a repeating participant. The Bayesian regression model used for this anslysis has the following form,
\begin{align}\label{eq:stage-2-varselect}
    Y^1_i \sim \text{Poisson}(\lambda_i), \quad
    \log(\lambda_i) = \hat\beta_0 + \bm{u}^\top_i \hat{\alpha} + \bm{v}^\top_i \hat{\beta}+ \bm{w}^\top_i \gamma, \quad \text{for } i = 1,2,\ldots,n_1
\end{align}
where $\tilde{\bm{v}}^\top_i$ is a vector containing only the selected features and $\hat\beta_0, \hat\alpha, \hat\beta$ are the posterior medians of a model of form  Eq. (\ref{eq:varselect-first-time}) fitted using standard normal priors for the coefficients of $\tilde{\bm{v}}_i$ as opposed to RHS priors. $\bm{w}_i$ contains the $33 = 14 + 2 + 5 + 9 + 3$ features of interest. As fatigue should only have negative effects on contact intensity, each element in $\gamma = (\gamma_1, \gamma_2, \ldots, \gamma_33)^\top$ is assigned a truncated RHS prior such that the values are constrained to the negative real domain. This was done by replacing  Eq.~\ref{eq:varselect-first-time} with a half-normal prior such that $\gamma_l | \zeta_l, \varepsilon, c \sim \text{half-Normal}^{-}(0, (1-2/\pi)^{-1} \varepsilon^2 \tilde{\zeta}^2_l)$ for $l = 1,2,\ldots,L$. For simplicity, we denote the constrained RHS prior as
\begin{equation}
    \gamma_l \sim \text{half-RHS}^{-}_{\nu_1,\nu_2,\nu_3}(s^2, p_0).
\end{equation}
The adjustment factor $(1-2/\pi)^{-1}$ in the scale component of the half normal was added to ensure that the prior variance of the constrained RHS prior is equivalent to that of the non-constrained version (for details, refer to Supplemental Text S1). As in the previous analysis, we set the hyperparameters of the RHS prior to $\nu_1 = 3, \nu_2 = 2, \nu_3 = 4, s^2 = 2$~\cite{piironen_sparsity_2017}, and $p_0 = L/2$. Features were selected if their median reduction effect on average contact intensity was greater than 5\%. We take the union of features obtained from applying the same model independently to survey wave 4 and 21 to obtain the final set.

\subsection*{Functional form of reporting fatigue effects over the number of repeat participations}
Previous studies showed that reporting fatigue build up over the number of times participants contribute to consecutive survey waves~\cite{dan_estimating_2023, wong_social_2023}. We leverage this empirical observation to identify a simple yet effective functional form for modeling longitudinal reporting fatigue. We focused on data from waves 3-12 (May 28\textsuperscript{th}, 2020 to July 1\textsuperscript{st}, 2020), as these waves fell into a period of homogeneous, relaxed contact restriction measures in Germany, thereby minimising the potential confounding effects of non-pharmaceutical interventions on our analysis. We also preferred to analyse data from relatively early period of the pandemic as interventions became more heterogeneous across states with time. We analysed a total 129,831 records from 2,492 unique participants with a maximum of 11 repeats.

Let $Y_{i,r}$ denote the number of contacts reported by individual $i$ in their $r$-th repeat for $i = 1,2,\ldots,N$ and $r = 0,1,\ldots,11$, where $r = 0$ indicates first-time participation. Let $t_{i,r}$ denote the calendar date in which participant $i$ had their $r$-th repeat. The model for longitudinal data has the following form:
\begin{equation}\label{eq:longitudinal-model}
    Y_{i,r} \sim \text{NegBinomial}(\lambda_i, \varphi), \quad
    \log(\lambda_i) = \beta_0 + \bm{x}^\top_i\beta + \tau(t_{i,r}) + \rho(r)
\end{equation}
where $\varphi$ denotes a real-valued overdispersion parameter such that the mean and variance of the Negative Binomial observation model are $\E{Y_{i,r}} = \lambda_{i,r}$ and $\Var{Y_{i,r}} = \lambda_{i,r} + \lambda_{i,r}^2/\varphi$. $\bm{x}_i$ is a vector features containing age, sex, household size, and other relevant features identified in the previous feature selection analysis. $\tau(t_{i,r})$ is models calendar-time effects as deviations from the baseline $\beta_0$ through a zero-mean Gaussian process characterised by the Mat\'ern 3/2 covariance kernel
\begin{equation}
k(t, t')  = \sigma_{\tau} \left(1 + \frac{\sqrt{3}|t - t'|}{\ell_\tau} \right) \exp\left( -\frac{\sqrt{3}|t - t'|}{\ell_\tau} \right).  
\end{equation}
Finally, $\rho(r)$ accounts for the reduction in overall contact reports as a function of the number of repeats $r$. We assessed 3 choices for modeling $\rho(r)$, which differed in the degree of assumption on the functional form of accumulating reporting fatigue. Throughout, we assumed that accumulating reporting fatigue effects did not change over calendar time to ensure identifiability of time trends in contact intensities, and additionally assumed that the strength and shape of reporting fatigue was identical across participants.

The first model comprised $R$ independent reporting fatigue effects where
\begin{equation}
    \rho_r \overset{i.i.d.}{\sim} \text{Normal}(0, 1) \quad r \in \{1,2,\ldots,R\},
\end{equation}
with $\rho_{r=0}$ set to $0$ for identifiability. This first model imposes no specific functional form of reporting fatigue with $r$. The second model was a non-parametric Gaussian process prior with squared exponential kernel depending on rescaled repeat times $\tilde{r} = (r - \text{mean})/\text{sd-dev}$
\begin{subequations}
\begin{align}
    \rho^{\text{GP}}(\tilde{r}) & \sim \text{GP}(0, k_{\text{SE}})\\
    k_{\text{SE}}(\tilde{r},\tilde{r}') & = \sigma_\rho \exp\left( -\frac{(\tilde{r}-\tilde{r}')^2}{2 \ell_\rho^2} \right) \\
    \sigma_{\rho}, \ell_{\rho} &\overset{i.i.d.}{\sim} \text{inv-Gamma}(5, 1)
\end{align}
\end{subequations}
for $r=1,\dotsc,R$, and $\rho^{\text{GP}}=0$ for $r=0$ to ensure identifiability. This model imposes smooth trends on reporting fatigue effects with the number of repeat participations, and consequently renders reporting fatigue effects more predictable.

For the third model, we take inspiration from dose-response modeling in biochemistry~\cite{prinz_hill_2009} and model reporting fatigue with a three-parameter Hill function:
\begin{equation}\label{eq:hill-function}
    \rho^{\text{Hill}}(r) = -\gamma \frac{e^\zeta r^\eta}{1 + e^\zeta r^\eta} \quad \text{where} \quad \gamma, \eta \in \mathbb{R}^+, \zeta \in \mathbb{R}
\end{equation}
where $\gamma$ is a scale parameter which control the effect size at large values of $r$, while $\zeta$ and $\eta$ in tandem controls how quickly the function approaches its minimum value. Figure S3 shows the functional form of the Hill function under different parameter values. This model is computationally more efficient than GP models, and renders due to its parametric form reporting fatigue effects strongly predictable. Given their constraints, we assign the following priors to each parameter:
\begin{align*}
    \gamma &\sim \text{half-Normal}^+(0,1) \\
    \zeta &\sim \text{Normal}(0,1) \\
    \eta &\sim \text{Exponential}(1).
\end{align*}
Throughout, we used the following generic priors to the parameters in the components common across all models:
\begin{subequations}
\begin{align}
1/\varphi &\sim \text{Exponential}(1), \\
\beta_0 &\sim \text{Normal}(0, 10), \\
\beta_j &\overset{i.i.d}{\sim} \text{Normal}(0, \sigma^2_{\beta}) \quad j = 1,\ldots,J \\
\sigma_{\beta} &\sim \text{Cauchy}^+(0,1) \\
\sigma_{\tau}, \ell_{\tau} &\overset{i.i.d.}{\sim} \text{inv-Gamma}(5, 1).
\end{align}
\end{subequations}

\subsection*{Correcting for long-term reporting fatigue effects with statistical contact models}
Contact intensity estimates gained from statistical models fitted to data with only first-time participants avoids the bias induced by reporting fatigue. However, this may mean that estimates are based on substantially smaller sample sizes (Figure~\ref{fig:1}). Here, we describe our approach to evaluate our proposed method for mitigating the confounding effects of reporting fatigue. The analysis is conducted on wave 21 which contains a large number of newly recruited participants as well as repeat participants with varying numbers of repeats. 

Let $a_i$ denote the age of participant, $\bm{u}_i = (u_{i1}, \ldots, u_{iP})^\top \in \{0,1\}^P$ denote the vector of sex and household size variables, and $\bm{w}_i = (w_{i1}, \ldots, w_{iQ})^\top \in \{0,1\}^Q$ denote the vector of covariates associated with reporting fatigue identified in the variable selection analysis. We consider the following negative binomial generalised additive count model,
\begin{subequations} \label{eq:gam}
\begin{align}
    Y_{i,r} &\sim \text{NegBinomial}\left( \lambda_{i,r}, \varphi \right), \\
    \log(\lambda_{i,r}) &= \beta_0 + \bm{u}_i^\top \beta + f_\theta(a_i) + \bm{w}^\top_i \rho(r)
\end{align}
\end{subequations}
where $\beta_0$ is a global baseline parameter and $\rho(r) \in \mathbb{R}_{\leq0}^Q$ is a vector consisting of elements
\begin{align*}
    [\rho(r)]_q = -\gamma_q \exp(\zeta_q)r^{\eta_q}/(1+\exp(\zeta_q)r^{\eta_q})
\end{align*}
which models the fatigue effect at each repeat for the selected variables.
$\alpha$ and the covariate parameter vector $\beta = (\beta_1, \beta_2, \ldots, \beta_p)^\top$ are assigned generic Gaussian priors, and the reciprocal of the overdispersion parameter $\varphi$ is given an exponential prior:
\begin{align*}
    \beta_0 &\sim \text{Normal}(0, 10), \\
    \beta_p &\overset{i.i.d.}{\sim} \text{Normal}(0, 1), \quad p = 1,\ldots,P \\
    1/\varphi &\sim \text{Exp}(1).
\end{align*}
Let $\hat\gamma, \hat\zeta$, and $\hat\eta$ denote the posterior median estimates from Eq.~\eqref{eq:longitudinal-model}. Leveraging the Bayesian framework to reflect prior knowledge, we assign the following informative priors to the parameters of the hill function,
\begin{align*}
    \gamma_q &\overset{i.i.d.}{\sim} \text{half-Normal}^+(\hat\gamma, 0.5), \\
    \zeta_q &\overset{i.i.d.}{\sim} \text{Normal}(\hat\zeta, 0.1), \\
    \eta_q &\overset{i.i.d.}{\sim} \text{half-Normal}^+(\hat\eta, 0.1).
\end{align*}
$f_\theta(a_i): \mathbb{R} \mapsto \mathbb{R}$ is a non-linear function that describes the effect of participant age on the rate of contact. For computational efficiency, $f_\theta$ is modeled using the zero-mean Hilbert space approximate Gaussian process~\cite{solin_hilbert_2020, riutort-mayol_practical_2022} (HSGP) which is parametrised by $\theta = (\sigma, \ell)$ and characterised by the squared exponential covariance kernel:
\begin{equation*}
    k(x, x') = \sigma \exp\left( -\frac{(x - x')^2}{2\ell^2} \right).
\end{equation*}

We begin by fitting the model to the subset of data containing only first-time participants and treat the medium estimates as the baseline. We then include participants with a single repeat such that the sample include participants with up to a single repeat. We proceed in this fashion until all participants is included. At each step, we evaluate the mean absolute percentage error (MAPE) against the baseline and the percentage of points in the baseline contained in the 95\% posterior credible intervals. We then applied the model to all 33 waves of the COVIMOD study and obtained population-level average contact intensity estimates weighted by age, sex, and household size. These estimates were compared against those obtained via a simple bootstrap procedure, model estimates without reporting-fatigue adjustment term $\bm{w}^\top_i \rho(r)$, and estimates from data on first-time participants only for waves with more than 300 first-time participants.

To obtain fatigue adjusted estimates for all 33 waves of the COVIMOD surveys, we applied the generalised additive model described above in the following manner. Let $t = 1,2,\ldots,33$ be an index for the waves. For the first wave ($t = 1$), priors were unchanged from those defined in the previous section. From the second wave onwards, priors for the parameters of the individual covariates and hill functions were set as,
\begin{align*}
    \beta_{t,p} \sim \text{Normal}(\hat\beta_{t-1,p}, 0.3) \quad \text{for } p = 1,2,\ldots,P
\end{align*}
and
\begin{align*}
    \gamma_{t,q} &\sim \text{half-Normal}^+(\hat\gamma_{t-1,q}, 0.3) \\
    \zeta_{t,q} &\sim \text{Normal}(\hat\zeta_{t-1,q}, 0.1) \\
    \eta_{t,q} &\sim \text{half-Normal}^+(\hat\eta_{t-1,q}, 0.1) \quad \text{for } q = 1,2,\ldots,Q.
\end{align*}
Here, $\hat\beta_{t-1,p}, \hat\gamma_{t-1,q}, \hat\zeta_{t-1,q}$, and $\hat\eta_{t-1,q}$ denote the posterior mean of the parameters from the previous wave. This sequential fitting process allows the model to partially leverage information from previous waves to resolve an un-identifiability issue between $\beta_{t,p}$ and the parameters of the Hill function when there are no first-time participants in the data. This issue does not exist for the non-fatigue adjusted Bayesian model, hence non-adjusted estimates were obtained by fitting the model independently to data from each wave.

%% file: supp_methods_1.tex
\section*{Supplementary Text S1}
The regularised horseshoe (RHS) prior\cite{piironen_sparsity_2017} defined as 
\begin{align*}
    \beta | \zeta_p, \varepsilon, c &\sim \text{Normal}(0, \varepsilon^2 \tilde{\zeta}^2_p), \ \tilde{\zeta}^2_p = \frac{c^2\zeta^2_p}{c^2 + \varepsilon^2\zeta^2_p} \\
    \zeta_p &\sim \text{student-}t^+_{\nu_1}(0, 1), \quad p = 1, 2, \ldots, P \\
    c^2 &\sim \text{inverse-Gamma}(\nu_2, \nu_2 s^2/2) \\
    \varepsilon &\sim \text{student-}t^+_{\nu_3}(0, \varepsilon_0),
\end{align*}
has the conditional mean $\E{\beta_p | \varepsilon, \tilde\zeta_p} = 0$ and variance $\Var{\beta_p | \varepsilon, \tilde\zeta_p} = \varepsilon^2 \tilde{\zeta}^2_p$. To constrain the horseshoe prior to the positive real domain $\mathbb{R}^+$, we can introduce an auxiliary random variable $z_p \sim \text{half-Normal}^+(0, \sigma^2)$ ($\E{z} = \sigma\sqrt{2/\pi}$, $\Var{z} = \sigma^2(1 - 2/\pi)$) and parameterise the RHS variable as $\gamma_p = \varepsilon \tilde{\zeta}_p \times z_p$. When $\sigma^2 = 1$, we can easily verify that the random variable $\gamma_p$ has conditional mean and variance,
\begin{equation*}
    \E{\gamma_p | \varepsilon, \tilde\zeta_p} = \sqrt{\frac{2}{\pi}} \varepsilon \tilde\zeta_p, \quad
    \Var{\gamma_p | \varepsilon, \tilde\zeta_p} = \left(1 - \frac{2}{\pi}\right) \varepsilon^2 \tilde\zeta^2_p \approx 0.36 \Var{\beta_p | \varepsilon, \tilde\zeta_p}.
\end{equation*}
Although the distribution has a non-zero mean, $\E{\gamma_p | \varepsilon, \tilde\zeta_p} \rightarrow 0$ as $\varepsilon\tilde\zeta_p \rightarrow 0$, ,{\it i.e.}, $\gamma_p$ is pulled towards zero by the global shrinkage parameter $\varepsilon$. To ensure that the variance of the constrained parameters $\gamma_p$ remain the same as the RHS prior, we can set the variance of the auxiliary random variable $z$ to $\sigma^2 = (1 - 2/\pi)^{-1}$ which gives us,
\begin{equation*}
    \Var{\gamma_p | \varepsilon, \tilde\zeta_p} = \left(1 - \frac{2}{\pi}\right)^{-1}\left(1 - \frac{2}{\pi}\right) \varepsilon^2 \tilde\zeta^2_p = \varepsilon^2 \tilde\zeta^2_p = \Var{\beta_p | \varepsilon, \tilde\zeta_p}.
\end{equation*}

%% file: supp_figure_1.tex
\renewcommand{\thefigure}{S\arabic{figure}}
\setcounter{figure}{0}
\begin{figure}
    \centering
    \includegraphics[width=\linewidth]{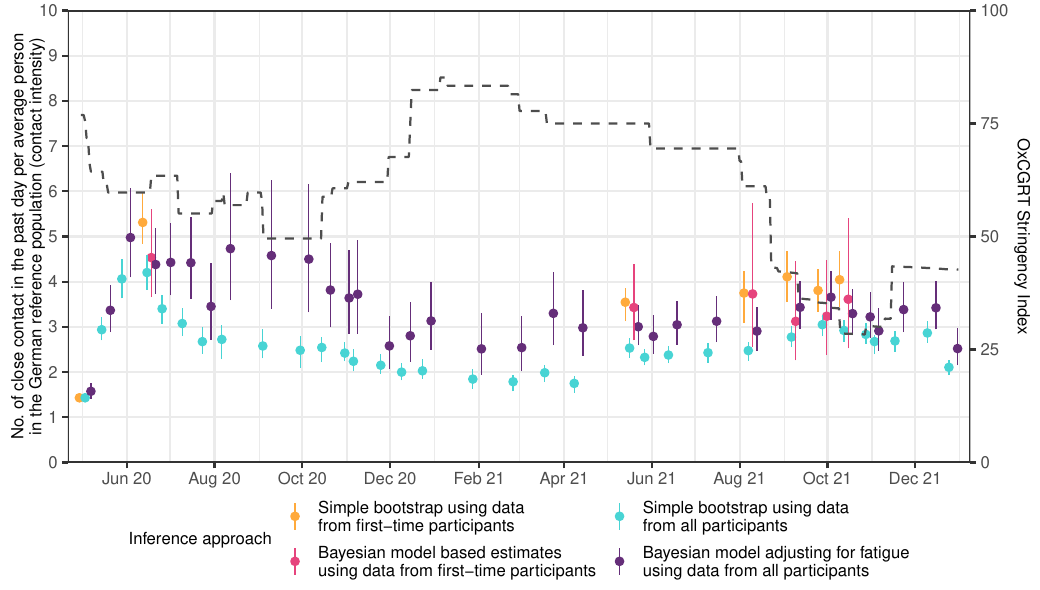}
    \caption{\textbf{Comparison between simple bootstrap based and Bayesian model based estimates for longitudinal contact intensity during the COVID-19 pandemic in Germany.}, national-level contact intensity estimates (point: simple bootstrap mean or posterior median estimate, linerange: 95\% bootstrap confidence or 95\% credible intervals) are shown according to different estimation approaches: Simple bootstrap~\cite{funk_socialmixr_2024} using data from first-time participants only, for waves with more than 300 first-time participants (orange); Simple bootstrap~\cite{funk_socialmixr_2024} using data from all participants and not adjusting for reporting fatigue (blue);
    Bayesian model using data from first-time participants only, for waves with more than 300 first-time participants (pink); Bayesian model using data from all participants and adjusting for reporting fatigue (purple). The dashed line represents the OxCGRT Stringency Index with higher values indicating a higher degree of contact restrictions (min: 0, max: 100).}
    \label{fig:sup1}
\end{figure}

%% file: supp_figure_2.tex
\begin{figure}
    \centering
    \includegraphics[width=\linewidth]{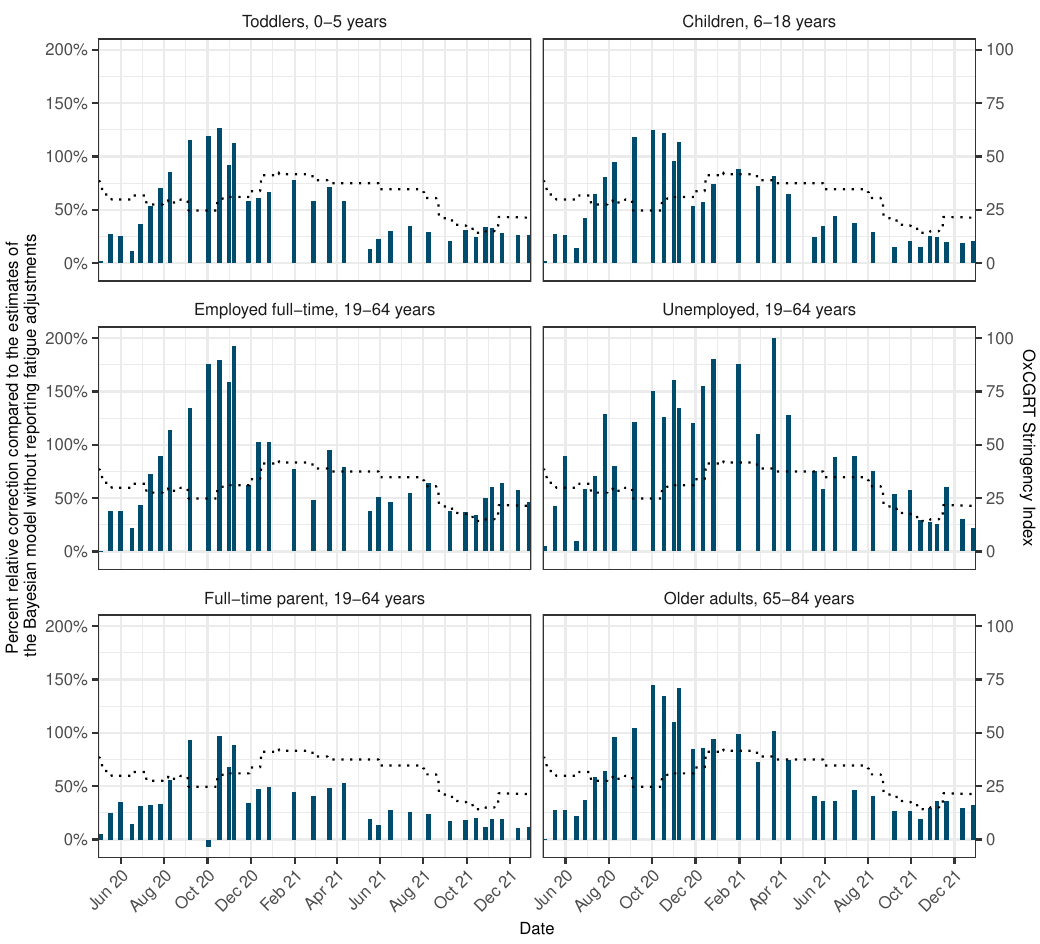}
    \caption{\textbf{Percent relative correction of contact intensity estimates from the Hill model against the Bayesian model without reporting fatigue adjustments}. Blue bars represent percent change in median contact intensity estimates from the Hill model relative to estimates from the Bayesian fatigue un-adjusted model. The dotted lines represents the OxCGRT Stringency Index with higher values indicating a higher degree of contact restrictions (min: 0, max: 100).}
    \label{fig:sup2}
\end{figure}

%% file: supp_figure_3.tex
\begin{figure}
    \centering
    \includegraphics{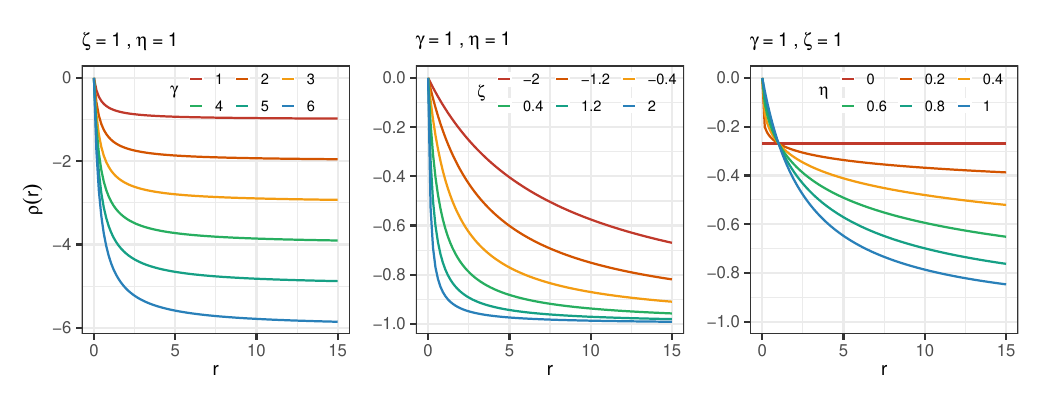}
    \caption{\textbf{Functional form of the Hill function under different parameter values.} Left: varying the scale parameter $\gamma$ from 1 to 6 while fixing shape parameters $\zeta$ and $\eta$ at 1. Centre: varying the shape parameter $\zeta$ from -2 to 2 while fixing the scale parameter $\gamma$ and the second shape parameter $\eta$ at 1. Right: varying the second shape parameter $\eta$ from 0 to 1 while fixing the scale parameter $\gamma$ and the first shape parameter $\zeta$ at 1.}
    \label{fig:hill-function}
\end{figure}